\documentstyle[aps,epsf,twocolumn]{revtex}

\begin{document}
\title{The Quantum Zeno Effect $-$ Evolution of an Atom Impeded by Measurement}
\author{Chr. Balzer, R. Huesmann, W. Neuhauser, and P.E. Toschek}
\address{Institut f\"{u}r Laser-Physik, Universit\"{a}t Hamburg, Jungiusstrasse 9,\\
D-20355 Hamburg, Germany\\
Tel.: 49-40-42838-2381, Fax: 49-40-42838-6571\\
E-mail: toschek@physnet.uni-hamburg.de}
\maketitle

\begin{abstract}
{\large \ }The evolution of a quantum system is supposed to be impeded by
measurement of an involved observable. This effect has been proven
indistinguishable from the effect of dephasing the system's wave function,
except in an individual quantum system. The coherent dynamics, on an optical 
$E2$\ line, of a single trapped ion driven by light of negligible phase
drift has been alternated with interrogations of the internal ion state.
Retardation of the ion's nutation, equivalent to the quantum Zeno effect, is
demonstrated in the statistics of sequences of probe-light scattering ''on''
and ''off'' detections, the latter representing back-action-free measurement.
\end{abstract}

\bigskip

\narrowtext The act of measuring an observable of a system that obeys
quantum mechanics consists of recording one of the eigenvalues and rejecting
all the other ones. This act is accompanied by sudden transition of the
system's wave function into the eigenfunction corresponding to the recorded
eigenvalue; the response of the system is known as the ''state reduction''
[1]. It has been recognized that repeated measurements retard, or even
impede, the evolution of a quantum system to the extent that they may
inhibit the evolution [2,3]. This consequence, the ''quantum Zeno effect''
[4] alluding to eleatic ontology, has aroused a great wealth of work devoted
to contemplating the subject [5], and an attempt to observe it: An
experiment including the drive and probe laser irradiation of an ensemble of
some 5000 beryllium ions confined in an ion trap has resulted in complete
agreement with quantum-mechanical predictions [6]. However, these
predictions based on the deletion, in the acts of measurement, of all
superpositions of eigenstates can be identified with the effect of any phase
perturbations by the environment upon the multi-particle wave function of
the system (''dephasing''). In fact, a perturbation via the back action of
the meter on the quantum system has been invoked as the origin of QZE
[7-10]. The ambiguity of the initial $t^{2}$evolution being set back by the
measurements [2,3,11], or dephasing, $-$ i.e. effect of measurement vs
dynamical effect $-$ is unresolvable since a decision would require
knowledge of the states of all the members (the ''micro-state'') of any
ensemble that remain unknown in a global measurement. Here, both the result
of a particular measurement, and the temporal evolution of the ensemble's
state, do not statistically depend on the results of previous measurements;
they are deterministic, save the ''projection noise'' [12,13] that affects
measurements of non-commuting observables and vanishes with a large enough
ensemble. However, with an individual system, the result of a measurement as
well as the system's evolution do statistically depend on the history, and
the results are in general found indeterministic, except after particular
preparation of the system, in an eigenstate of the observable to be detected
[13]. The statistics of the results will embody the signature of the state
reductions by the measurements, and their effect cannot be ascribed to
dephasing [14, 15]. This argument has been quantified, by numerical
calculation, for a single spin-like quantum system interacting with a light
mode whose photon number is measured, and for a corresponding ensemble [16]:
With a single quantum system, the evolution is not revealed by reiterated
measurements. The state of this system is reduced to an eigenstate, by each
precise enough measurement, according to the result of the measurement. As
long as the evolution of the quantum system between two subsequent
reductions is coherent, there is no base for invoking dephasing. This is so
because the configuration space of, say, an individual spin system extends
only over the surface of the unit sphere in SU2 symmetry, and the
micro-state of this system is completely known from the result of a
measurement, in contrast with that of an ensemble of spins.

In the experiment of [6], the state of the system has been interrogated only
after sequences of ''measurement'' pulses, such that back-and-forth
transitions go unnoticed and falsify the probability of survival [17]. In
contrast, here the result of each measurement is registered.

We have demonstrated the retardation of the light-driven quantum evolution
of an individual, localized cold ion upon repeated reduction by intermittent
probing the ion's two energy eigenstates involved in the driven resonance. A
single ion, $^{172}Yb^{+},$\ was localized in the node of the electric field
of a 2-mm-sized electrodynamic trap in ultra-high vacuum [18].\thinspace The
ion was centred in the pseudo-potential of the trap with less than 30nm
error, and laser-cooled to the Doppler limit (mean vibrational state $%
\langle n\rangle $\ $\simeq $\ $10$) in order to minimize the driven
micro-motion, and the free harmonic secular motion, respectively. The $E2$\
transition $S_{1/2}-D_{5/2}$\ was coherently driven, during time intervals $%
\tau =2ms,$\ by shining upon the ion 411-nm quasi-monochromatic light
generated by frequency-doubling the light of an 822-nm diode laser (Fig. 1).
The spectral width of the blue output did not exceed 500 Hz, with 1s of
integration.\thinspace The pulses of this drive light alternated with pulses
of 369-nm probe light, generated by frequency-doubling the output of a cw
LD700 dye laser.\thinspace The probe light excites some 10$^{8}$\ events/s
of resonance scattering on the $S_{1/2}-P_{1/2}$\ line of the ion only if
the ion is found in the $S_{1/2}$\ ground state after the driving pulse has
been applied [19].\thinspace The probe pulses were 10-ms long such that
clear-cut presence or absence of resonance scattering represent measurements
of the ion's internal state: Scattered light proves the ion to reside in the
ground state immediately after the detection of each photon, the absence of
light scattering instead reduces the ion to the $D_{5/2}$\ state.

\begin{figure}[tb]
\epsfxsize8cm
\epsffile{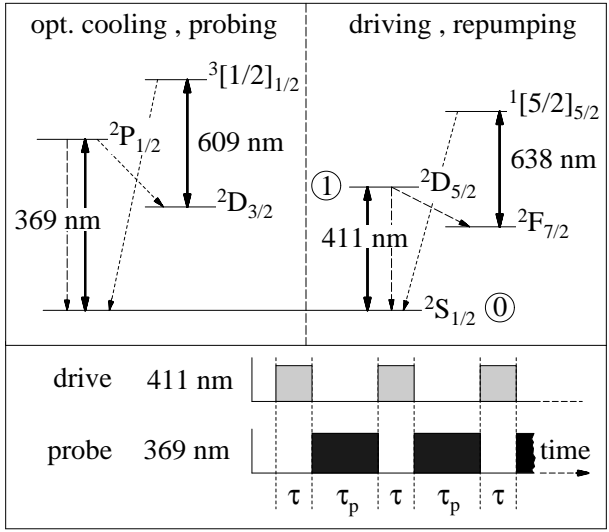}
\vspace{-2.3cm}
\caption{Energy levels and active transitions of $^{172}Yb^{+}$\ ion, and
program of alternating drive and probe light. $\protect\tau =2ms,$ $\protect%
\tau _{p}=10ms$.}
\end{figure}
The decay of the ion's $D_{5/2}$\ state into the metastable state $F_{7/2}$\
of extreme lifetime [20] complicates the dynamics on the driven $E2$\ line.
However, we kept the ion continuously irradiated by the cw output light of a
diode laser at 638nm that completely saturates the $F_{7/2}-$\ $%
^2[5/2]_{5/2} $\ line of the ion in order to immediately repump the ion from
the $F_{7/2}$\ level into its $S_{1/2}$\ ground state.

A measurement based on light scattering absent (''off''), i.e., when the ion
is in its $D_{5/2}$\ state, extends no physical reaction on the ''quantum
object'' (i.e., the spin represented by the $E2$\ transition), it is of the
quantum non-demolition type [10]: (i) Both the quantum object and the
''quantum probe'' (the dipole on the resonance line) return to their initial
states after the measurement. (ii) The probe light does neither cause any
dissipation nor back action in the combination of quantum object and quantum
probe. The state of this probe is indirectly measured by the null detection
of scattered light, with zero recoil upon the ion. This state is correlated
with the upper one of the two alternative eigenvalues of the observable to
be measured, the internal energy of the quantum object.

Probe-light scattering makes the ion recoil. Its net effect is spatial and
velocity fluctuations, and a random phase shift upon the quantum object, the
driven quadrupole. However, the ion remains cooled deep inside the
Lamb-Dicke regime (excursion $\ll $ light wavelength), and the corresponding
phase variations do not exceed a small fraction of $\pi $.

The condition for strong trapping [21] holds with the driven $E2$\ line,
such that any net recoil of the drive light is indeed absorbed by the trap
since the ion's vibrational frequency, 1.3 MHz, far exceeds the natural line
width.

The crucial issue is the capability, of the two kinds of measurements, to
distinguish retarded evolution from the effect of dephasing. The phase of
the drive light is found to diffuse, in each 2-ms interval of irradiation,
by such a small fraction of $\pi $\ only, that there is no risk of quenching
the coherence. The ion's super-position state would be conserved during the
breaks of the drive, as long as it is not measured, for t $<$\ $\Gamma ^{-1}$%
\ $=$\ the decay time of the inversion$.$\ State reduction by each
subsequent probing yields random results, and the coherent dynamics is
retrieved only when averaging over an ensemble of measurements, after
identical preparation. $-$\ Preservation of the ion's\ vibronic state under
probing the states of an electronic resonance has permitted detection of the
corresponding nutational dynamics by a stroboscopic or ''sampling''
technique [22].

Fig. 2\ shows a trajectory of ''on'' and ''off'' events. The number of
on/off pairs accumulated over $500$\ measurements and normalized by the
number of ''on'' events yields the probability of excitation, on the $E2$\
line, to the metastable $D_{5/2}$\ level. With negligible relaxation, this
transition probability is 
\begin{equation}
p_{01}={\it \cos }^{2}\chi \cdot {\it \sin }^{2}(\theta /2),{\large \ } 
\eqnum{1}
\end{equation}
\noindent where tan$\chi =\Delta /\Omega $, $\theta =\sqrt{\Omega
^{2}+\Delta ^{2}}$\ $\tau $, and $\Omega $\ and $\Delta =\omega -\omega _{0}$%
\ are Rabi frequency and detuning of the driving light $\left( \omega
\right) $\ off its resonance $\omega _{0}$, respectively.
\begin{figure}[tbp]
\epsfxsize8cm
\epsffile{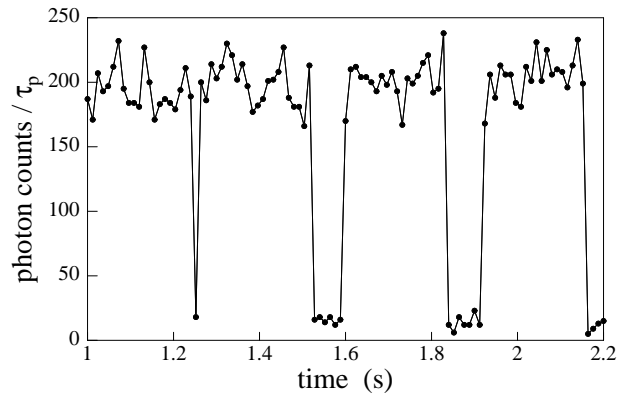}
\vspace{0.5cm}
\caption{Part of trajectory of results of measurements each of which
consists of a drive and a probe pulse applied to the ion.}
\end{figure}

When this data acquisition is repeated either at varied length $\tau $\ of
the drive pulse, or at stepwise progressive detuning of the drive frequency $%
\omega $\ ,\ the corresponding mean rate oscillates as a result of sampling
the effective Rabi nutation frequency $\theta /\tau $\ (Fig. 3). The
observation of this modulated excitation spectrum proves the coherent nature
of the interaction of ion and drive light. Note the first-order sideband, at
1.3 MHz up-tuning, generated by residual vibrational phase modulation of the
ionic quadrupole.

Let us turn to the statistics of trajectories, as in ${\rm Fig.\,2}$. When
starting with the ion in the ground state, another ''on'' event takes place
with probability $p_0={\it \cos }^2$($\Omega \tau /2)$. When starting with
the ion in the metastable state, an ''off'' result would take place with
same probability, $p_1=p_0$, if we neglect relaxation, for the moment. In
each driving pulse, the light-driven nutation starts anew thanks to the
state reduction by the previous probing, and it extends over the pulse
duration $\tau $. The next probe pulse reduces the ion again to one of the
two energy eigenstates. Then, the probability of finding either ''on'' $%
(i=0) $\ or ''off'' $(i=1)$\ \ $q$\ times in a sequence, is 
\begin{equation}
U(q)=U(1)V(q-1)  \eqnum{2}
\end{equation}
\noindent where 
\begin{equation}
V(q)=p^q={\large cos}^{2q}(\Omega \tau /2).  \eqnum{3}
\end{equation}
In contrast, with state reduction lacking the ion's evolution would continue
coherently over the total time of driving, as long as the laser-induced
quadrupole moment survives ($q\tau <\gamma ^{-1}=$ time of residual
dephasing). This dynamics would require $V(q)=$\ cos$^2(q\Omega \tau /2)$.

Actually the ion dynamics is modified by relaxation: The two probabilities $%
p_i$\ must include (i) spontaneous decay, and (ii)\ phase (''transverse'')
relaxation of the ionic quadrupole. The Bloch equations for a spin system
include these processes [23]. From an analytic solution on resonance ($%
\Delta =0$) [24] one derives 
\begin{equation}
\hspace{-2pt}p_i=1-\hspace{-2pt}f_iB_i\left( \hspace{-2pt}1-\hspace{-2pt}%
\sqrt{1+\tan ^2\epsilon _i}\text{ }e^{-\left( a+b\right) }\cos \left( \theta
-\epsilon _i\right) \right) \hspace{-2pt},  \eqnum{4}
\end{equation}
\noindent where $\tan \epsilon _1=(a-b-2b(\Omega \tau )^2/[(\Omega \tau
)^2+8\,ab])/\theta ,$\ $\tan \epsilon _0$\ $=\left( a+b\right) /\theta ,$\
and $B_0=\frac{\Omega ^2/2}{\Omega ^2+\Gamma \gamma },$\ $B_1=1-B_0,$\ $%
2a=\gamma \tau =\gamma _{ph}\tau +\left( \Gamma /2\right) \tau ,$\ $%
2b=\Gamma \tau ,$\ $\theta ^2=\left( \Omega \tau \right) ^2-\left(
a-b\right) ^2.$\ The supplementary factor $\ f_i$\ \ is unity for
non-degenerate levels, the net probability of excitation is $p_{01}$\ $=$\ $%
1-p_{0\text{ }}$, and $\gamma _{ph}$\ is the rate of phase diffusion, and $%
\Gamma /2$ is the decay rate of state $D_{5/2}$.

The probabilities $p_{0}$\ and $p_{1}$\ no longer agree. The rate $\gamma
_{ph}$\ results from residual phase fluctuations of the drive laser. A
simple model of phase diffusion [23] yields $\gamma _{ph}=D/2.$\ The
diffusion constant $D$\ is related to the phase variance as $\left\langle
(\delta \varphi )^{2}\right\rangle =D\tau $, such that the standard
deviation\thinspace of\thinspace the drive's phase\thinspace is $\overline{%
\delta \varphi }=\sqrt{\left\langle (\delta \varphi )^{2}\right\rangle }=%
\sqrt{2}$\thinspace $\sqrt{2a-b}$.
\vspace{-0.5cm}
\begin{figure}[tbp]
\epsfxsize8cm
\epsffile{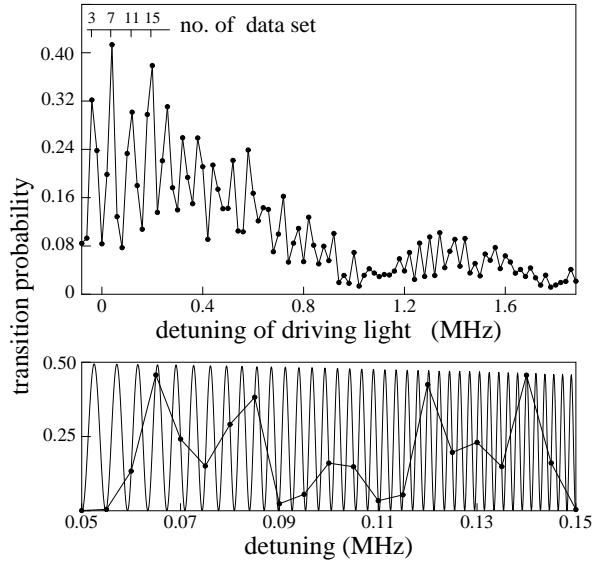}
\vspace{-1.5cm}
\caption{Probability of excitation, vs detuning $\protect\nu -\protect\nu
_{o}$\ by 20-kHz steps of the drive (top). Note the first-order vibronic
sideband at 1.3 MHz. Within a small range close to resonance, detuning
replaces variation of the drive-pulse length $\protect\tau $\ . The spectrum
of absorption is superimposed by stroboscopic sampling of the ion's Rabi
nutation, as demonstrated by a simulation with small steps on an expanded
scale (bottom).}
\end{figure}

The $S_{1/2}$\ ground state is resolved in two Zeeman-split sublevels one of
which only is excited by the drive light, leaving $f_0=1/2,$\ whereas the
less resolved upper state $^2D_{5/2}$\ suggests $f_1=1.$\ From a fit of $%
p_{01\text{ }}$to its contrast of modulation, i.e., to the extreme values of
the data sets 7 and 9 close to resonance in ${\rm Fig.\,3}$, one derives the
values $B_0=0.49$\ and $a+b\simeq $\ $0.38.$\ The approximate phase of
nutation achieved by the 2-ms-long driving pulse is derived from the
contrast on the wing of the power-broadened resonance (data sets no. 15 and
16) using eq. (1): $\theta _{app}\simeq \Omega \tau \cong 578\times 2\pi .$\
A better value is revealed by the increment of nutation per each step of
detuning, $\delta \omega =20kHz\times 2\pi ,$\ that is $\delta \theta =\sqrt{%
\theta ^2+(\tau \delta \omega )^2}$\ $-$\ $\theta =0.5\pi $\ ${\it 
%TCIMACRO{\func{mod}}%
%BeginExpansion
\mathop{\rm mod}%
%EndExpansion
}$\ $2\pi .$\ Compatible with $\theta _{app}$\ is only $\delta \theta
=1.25\times 2\pi ,$\ and $\theta =2\pi n+\theta ^{\prime }$,\ where the
integer $n\simeq 640,\ $and $\theta ^{\prime }\ $is a fraction of $2\pi $.

The numbers of sequences made up of $q$\ identical results have been
evaluated from each trajectory, and they are identified with $U(q)$. ${\rm %
Fig.\,4}$\ shows data of $U(q)/U(1)$, recorded at the indicated three
settings of the frequency detuning close to the principal resonance of ${\rm %
Fig.\,3}$. It shows also values of $V(q-1)$\ calculated from the solutions
of $p_{0}$\ and $p_{1},$\ and multiplied by the factor $(N-q+1)/N$\ that
takes care of the finite length of the data sets. From the fit to the ''on''
sequences of data set 7, $a+b$\ $=0.395$\ and $\theta ^{\prime
}=(1+10^{-4})\pi $\ are derived. The constant of total relaxation is
inserted when fitting the ''on'' statistics of the neighbouring data sets
for their fractional phases $\theta ^{\prime }$. The same phase values $%
\theta ^{\prime }$\ attributed to the ''off'' statistics require $f_{1}$\
decreasing from unity with increasing transition probability $p_{10}:$\
Cycles of spontaneous decay and reexcitation increasingly contribute to the
''off'' sequences and redistribute the ion over the $D_{5/2}$\ sublevels
such that potential deexcitation becomes selective, and $f_{1}$\ is expected
to approach $1/6$.
\vspace{-0.5cm}
\begin{figure}[tbp]
\epsfxsize8cm
\epsffile{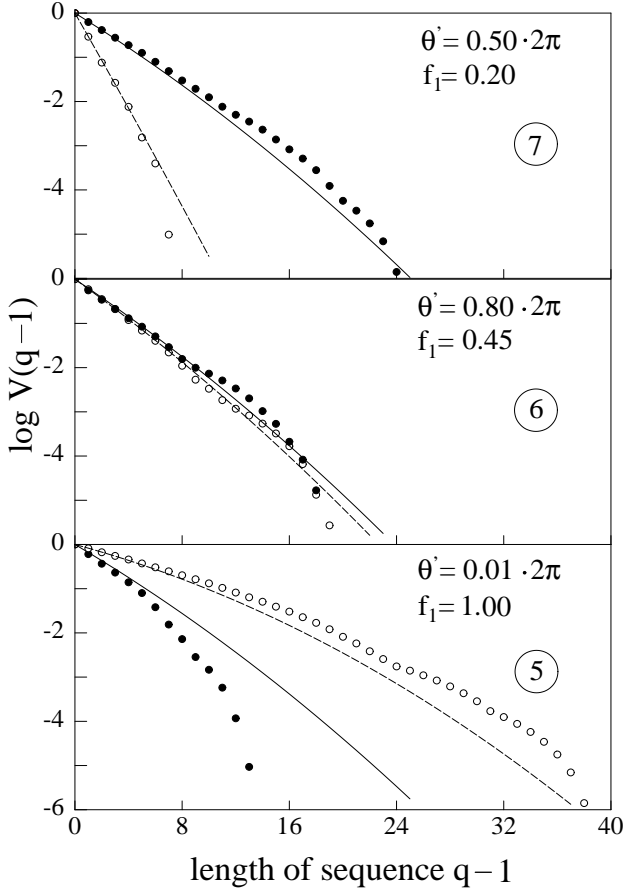}
\vspace{-3.5cm}
\caption{Probability $U(q)/U(1)$\ of uninterrupted sequences of ${\it q}$\
''on'' results (white dots) and ''off'' results (black dots). The lines show
the distributions of probabilities $V(q-1)$\ for the ion's evolution on its
drive transition, according to Eqs. 2 and 3. $\protect\theta ^{\prime }$\
and $f_{1}$\ from fit; values $f_{1}<1$\ indicate redistribution, over
sublevels, by cycles of spontaneous decay and reexcitation.}
\end{figure}

The agreement of $U(q)/U(1)$\ and $V(q-1)$\ confirms state reduction by each
measurement that goes along with probing the ion's resonance scattering, and
it proves the coherent evolution of the ion to set out again after each of
these state reductions. The statistics of ''no count'' sequences provides
such a proof even under QND conditions$.$\ The overall effect of the
reiterated resettings is the ion's quantum dynamics being impeded.

Since $\theta \gg 2\pi ,$\ the resonance of Fig.\thinspace 3\ is
substantially power-broadened, and the other natural mode of relaxation
given by\ $a-b$\ cannot be determined. However, $a+b$\ is compatible with
the lifetime of the $D_{5/2}$\ level (5,7ms [20], 7,2ms [25]). It serves
also as an upper limit of a: For the standard deviation of the driving
phase, the limit yields $\overline{\delta \varphi }<1.2$\ $\ll 2\pi $. This
phase fluctuation corresponds to the bandwidth less than 100Hz, which is
compatible with the controlled 1-s frequency bandwidth, $\delta \nu =\delta
\varphi /\tau \lesssim 500$\ Hz. This restriction secures the interaction of
light and ion being coherent during the pulse length $\tau $, and it proves
that not dephasing but the effect of state reduction in the act of
measurement is responsible for the observed values of the joint
probabilities $U(q)$. Moreover, the data demonstrate that an individual atom
interacts coherently on the drive transition. So far coherent manipulation
has been demonstrated with a Raman [26] or a vibronic [22] transition.

In conclusion, we have demonstrated the inhibition of the evolution of a
quantum system by repeated measurements on the system, i.e., the quantum
Zeno effect, by alternately driving and probing an individual $^{172}Yb^{+}$%
\ ion on a weak ($E2$) and a strong (resonance) transition, respectively. We
have revealed the statistics of uninterrupted sequences of observations
which find the ion either in the ground state subjected to resonance
scattering of the probe light, or in the $D_{5/2}$\ state that does not show
resonance scattering since it is decoupled from the probe resonance. The
evolution derived from this statistics agrees with the evolution of the
ion's wave function assumed interrupted by state reductions, and the
retardation cannot be attributed to dephasing. The frustrated attempts of
detecting fluorescence qualify as quantum non-demolition measurements of the
internal ion energy. The statistical distribution of sequences of such
measurements identifies the degree to which the measurements, intertwined
among the pulses of driving, have impeded the quantum evolution of the ion.

This work was supported by the K\"{o}rber-Stiftung, Hamburg, by the
Hamburgische Wissenschaftliche Stiftung, and by the ZEIT-Stiftung, Hamburg.

\end{document}